\documentclass[lettersize,journal]{IEEEtran}
\usepackage{amsmath,amsfonts}
\usepackage{algorithmic}
\usepackage{algorithm}
\usepackage{array}
\usepackage[caption=false,font=normalsize,labelfont=sf,textfont=sf]{subfig}
\usepackage{textcomp}
\usepackage{stfloats}
\usepackage{url}
\usepackage{verbatim}
\usepackage{graphicx}
\usepackage{cite}
\begin{document}

\title{A 0.03${mm}^2$ 100-250MHz Charge-Pump or Amplifier-Less Integrating
Sub-Sampling PLL for Ultra-low Power Communication and Computing}

\author{Yudhajit Ray, Archisman Ghosh, ~\IEEEmembership{Graduate Student Member,~IEEE}\\
Shreyas Sen,~\IEEEmembership{Senior Member,~IEEE}}



\maketitle

\begin{abstract}
Clock generation is an essential part of wireless or wireline communication modules. To facilitate recent advancements in wireline-like communication and in-sensor computing modules at relatively lower data rates, ultra-low power, and accurate clock generation are of the utmost importance. This paper presents a unique implementation of integrating sub-sampling phase locked loop, which alleviates the usage of additional gain elements in the PLL and reduces the noise injection in the system. In this design, the ring oscillator-based PLL can operate a wide frequency range of 100-250MHz while consuming 0.03${mm}^2$ of area and 131.8$\mu W$ of power at 250MHz. The area-normalized figure of merit (FOM) of the integrating SSPLL is found to be -236, while showing a reference spur of -43.2dB.
\end{abstract}

\begin{IEEEkeywords}
Integrating Sub-sampling Phase Detector (ISSPD), Integrating Sub-sampling Phase Locked Loop (ISSPLL), Frequency Synthesizer, Figure-of-Merit (FOM), Low Power
\end{IEEEkeywords}

\vspace{-1em}\section{Introduction}
\IEEEPARstart{R}{ecent} advancements in low-power camera, body nodes, and communication modules have brought a need of generating highly accurate clocks at a relatively lower frequency ranges. Low-jitter and low-power frequency synthesizers are primarily used in wireline or wireless communication interfaces, data converters, in-memory computation modules, hardware acceleration, and many other crucial integrated circuits. There has been significant progress in developing high-frequency PLLs (approx. 0.4-100 GHz), improving the phase noise (PN), reference spur, and power consumption of such designs. 

Recently, there have been significant advancements in wireline-like channels like proximity communication\cite{b1} or body communication\cite{b2},\cite{b3}, which operate at the frequency range of 10-500 MHz. These communication modules require ultra-low power clock generation to ensure slow depletion of limited power supply on-body nodes. For these applications, low power consumption takes priority over low jitter clock generation.

\begin{figure}
\centerline{\includegraphics[width=3.5in]{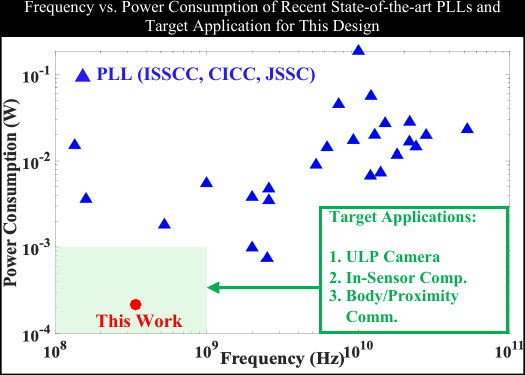}}
\caption{Frequency vs. Power Consumption of State-of-the-art PLL Implementations and Target Applications for this design \label{fig1}}
\end{figure}

Moreover, emerging video sensor nodes with in-sensor computing\cite{b4} require clock signals with high spectral purity while ensuring low power consumption for accurate data conversion at low power overhead. Because of these requirements, it is crucial to investigate low-power clock generation circuits at low to medium frequency of operation. Fig. \ref{fig1} shows the frequency of operation vs. power consumption comparison of recent state-of-the-art PLL implementations. There is a severe lack of innovations for low to medium-frequency (50-1000MHz) PLLs with ultra-low power and area consumption.   
\vspace{-1em}\section{Background and Relevant Work}
Sub-sampling PLLs provide a divider-less solution for generating low-jitter clock signals due to the inherent high phase detection gain ($K_{PD}$) of the sub-sampling phase detector. Moreover, the absence of a divider prohibits the addition of extra noise into the system, while avoiding a large multiplication factor for in-band phase noise. 
\begin{figure*}
\centerline{\includegraphics[width=7in]{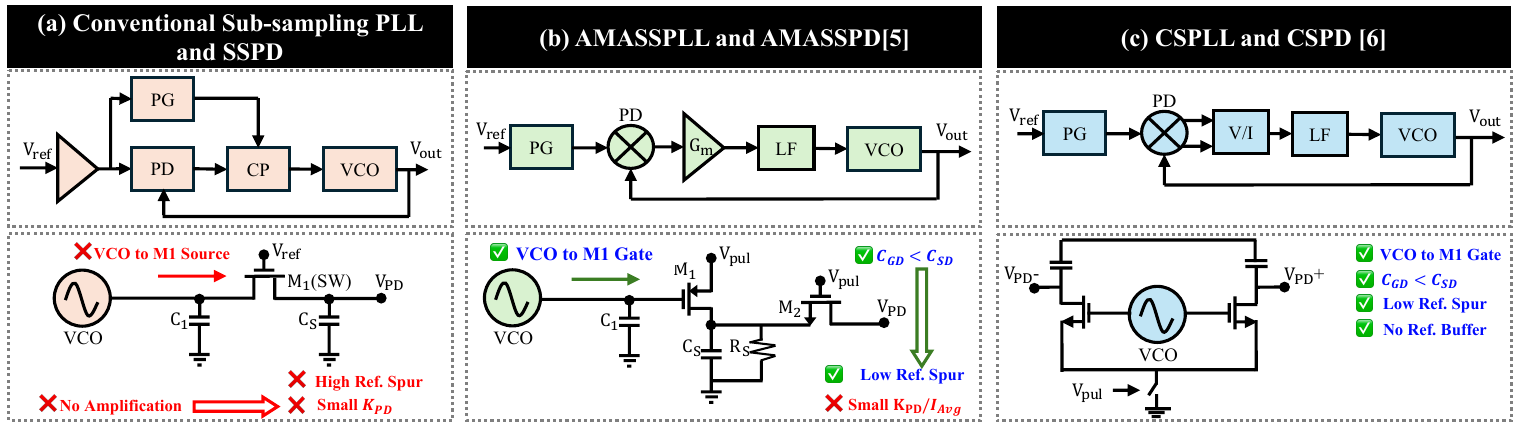}}
\caption{Sub-sampling PLL Architectures and Sub-sampling Phase Detector Architectures for (a) Conventional, (b) AMASSPLL \cite{b5}, and (c) CSPLL \cite{b6}
\label{fig2}}
\end{figure*}

However, sub-sampling PLLs suffer from high reference spur due to reference clock feed-through and charge injection through the sampling switch. Fig. \ref{fig2}(a) shows the overall architecture of conventional SSPLL along with conventional Sub-Sampling Phase Detector (SSPD). As the Voltage Control Oscillator (VCO) outputs are injected directly into the source of the sampling switch, the VCO tank observes a larger capacitance. Hence, an isolation buffer is often used to reduce the reference spur, increasing the power and area consumption of the SSPLL. 

\cite{b5} addresses this issue by using an active-mixer-adopted SSPD (AMASSPD) as shown in Fig. \ref{fig2}(b). In this architecture, the VCO outputs are fed into the gates of the SSPD, and it operates as a sample and hold circuit, controlled by the pulse signal. It shows a significant improvement in the reference spur. However, this phase detector does not provide high $K_{PD}$ and requires a transconductance amplifier stage, increasing the overall area. 

In \cite{b6}, a different approach has been proposed where a charge-sampling phase detection mechanism is used to provide high $K_{PD}$ in the Charge-Sampling Phase Detector (CSPD) while avoiding the need for isolation buffers. Large input transistors are necessary to provide sufficient gain. Moreover, a voltage-to-current converter is necessary to convert the differential voltage to current. 

\vspace{-1em}\section{INTEGRATING SUB-SAMPLING PLL}
\begin{figure}
\centerline{\includegraphics[width=3.5in]{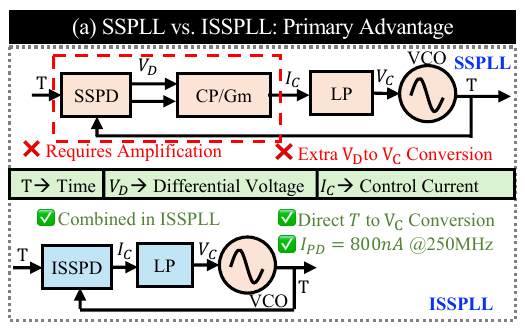}}
\caption{Primary advantage of ISSPD over other SSPD architectures\label{fig3}}
\end{figure}

The presence of a charge pump, trans-conductance amplifier, or voltage-to-current converter increases the amount of injected noise-current into the sub-sampling PLL while increasing the overall power and area consumption. The proposed architecture provides higher $K_{PD}$ in comparison to \cite{b5} and \cite{b6} without requiring the help of extra circuit components in the SSPLL, as shown in Fig. \ref{fig3}. 

The key contributions of this work are as follows: (1) This implementation alleviates the need for a charge pump, trans-conductance amplifier, or voltage-to-current converter in the sub-sampling PLLs. Thereby, reducing the amount of noise injected into the system. (2) The unique ISSPD gain response provides an opportunity to further reduce the in-band phase noise with an appropriate choice of pulse width ($T_{pul}$). (3) A 32-stage cross-coupled ring oscillator with RC degeneration for fine-tuning has been used to minimize the area and power consumption of the ISSPLL. At 250MHz the overall power consumption is 131.8$\mu W$, while showing an FOM of -236dB. 
\vspace{-2.5em}\subsection{Operating Principle}
\begin{figure*}
\centerline{\includegraphics[width=7in]{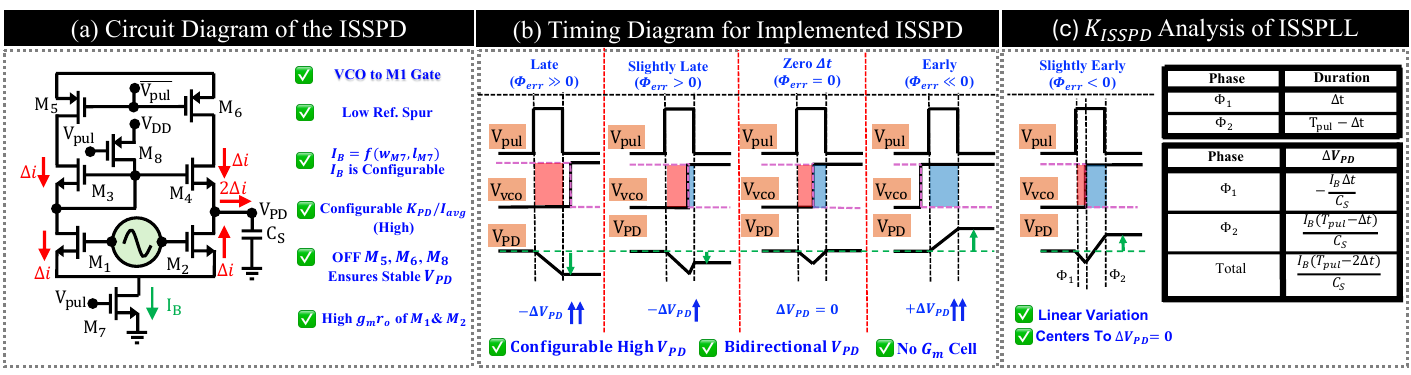}}
\caption{(a) Circuit diagram, (b) operating principle and timing diagram, and (c) phase detection gain analysis of the implemented ISSPD\label{fig4}}
\end{figure*}
The operating principle of the ISSPD (Fig. \ref{fig4}(a)) can be understood using the timing diagram as shown in Fig. \ref{fig4}(b). Depending on the position of the pulse ($V_{pul}$) with respect to the VCO transition, the integrated voltage on the $C_{s}$ changes. When the VCO transition is late with respect to the pulse ($\phi_{err}>0$), the integrated voltage ($\Delta V_{PD}$) becomes negative. An opposite effect can be observed when the VCO transition is early with respect to the pulse position. When the position of the VCO output transition is directly in the middle of the pulse, the effective change of PD output voltage ($\Delta V_{PD}$) stays zero. Hence, the ISSPLL loop can operate to gravitate the VCO transition towards the mid-point of the pulse. 
An advantage of using ISSPLL over CSPLL\cite{b2} implementation is the relaxed design restriction for PD architecture. This design prioritizes keeping $\Delta V_{PD}=0$, rather than forcing the transition to be at the mid-point, which requires ideal device sizing and, a strict locking scheme, and it becomes susceptible to process, voltage, or temperature variations. As a result, ISSPD architecture avoids additional design restrictions. For asymmetric device sizing in ISSPD, the ISSPLL will gravitate towards the time point where $\Delta V_{PD}=0$, which has a fixed reference clock period of $T_{REF}$.

Phase detection gain analysis of ISSPD ($K_{ISSPD}$) has been shown in Fig. \ref{fig4}(c). As it can be seen, the $\Delta V_{PD}$ is dependent on bias current ($I_{B}$), $T_{pul}$, phase difference ($\Delta t$), and $C_S$. In ideal scenario, when $\Phi_1$ and $\Phi_2$ phases have same current flow at $C_S$, $\Delta V_{PD}=0$ when $T_{pul}=2*\Delta t$. 
\begin{figure}
\centerline{\includegraphics[width=3.5in]{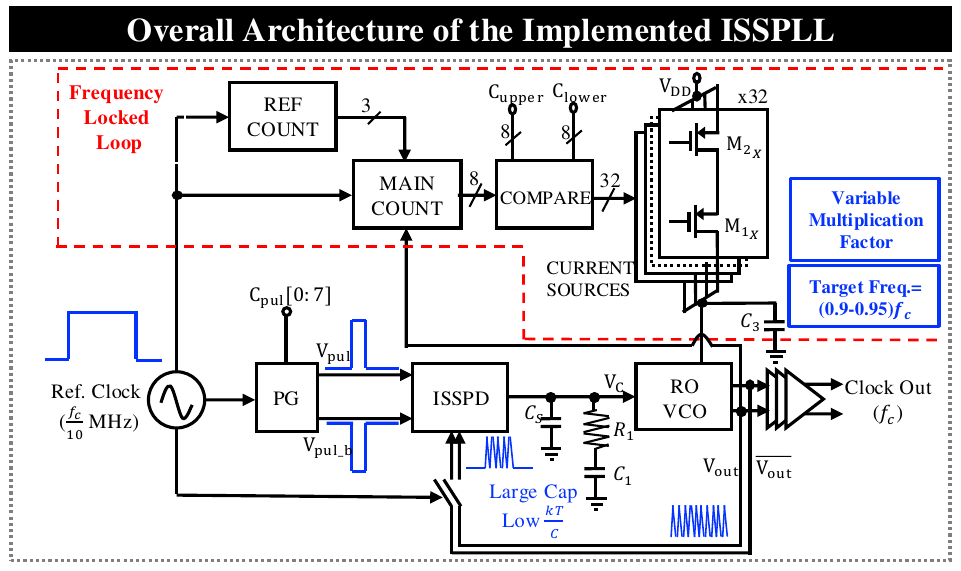}}
\caption{Overall Architecture of Integrating Sub-sampling PLL\label{fig5}}
\end{figure}
\begin{figure*}
\centerline{\includegraphics[width=7in]{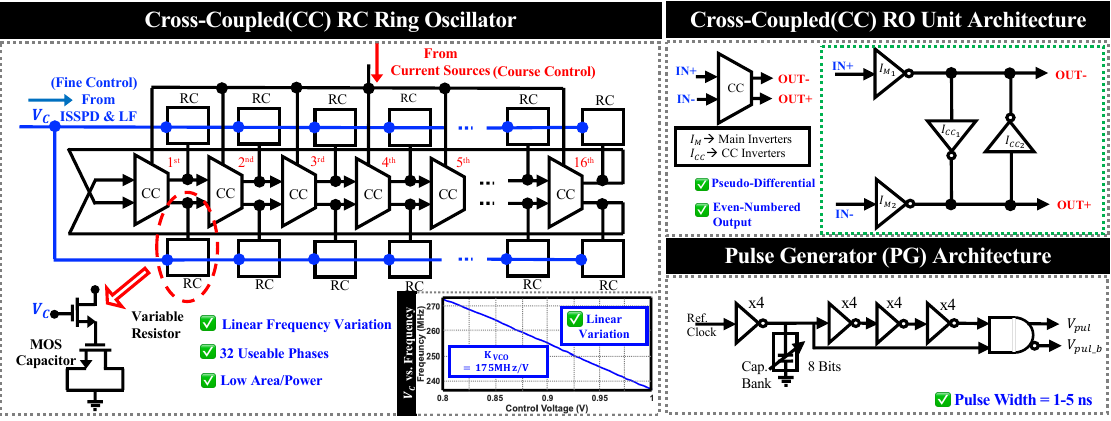}}
\caption{Cross-coupled RC Ring Oscillator with 32 Equally-Spaced Phases, pseudo-differential cross-coupled unit, and pulse generator architecture\label{fig6}}
\end{figure*}
\vspace{-1em}\subsection{Overall Architecture}
Fig. \ref{fig5} shows the overall architecture of the implemented integrating sub-sampling PLL. It comprises of a course frequency locked loop (FLL) and fine integrating sub-sampling PLL loop. All the sub-systems are as follows:
\subsubsection{Integrating Sub-sampling Phase Detector}
The implemented circuit diagram of the ISSPD is shown in Fig. \ref{fig4}(a), which follows a single-ended integrating amplifier architecture\cite{b7}. The differential current($\Delta i$) between M1 and M2 depends on the polarity of the VCO outputs, and the current flowing to or from the load capacitor ($C_S$) is double the differential current ($2\Delta i$) because of the source connected transistor $M_3$. As the bias current is a function of $M_7$'s device sizing, the gain can be configured as per the requirements, supporting a larger frequency range of operation. 

As the differential current is integrated over the load capacitor, it can be combined with the loop filter to generate the fine control voltage for the VCO. As a result, the loop filter area consumption is also optimized. 
\subsubsection{Loop Filter}
ISSPD along with $C_S$ produces a single pole at zero due to the integration property. As a result, just a series resistance (R1) and capacitance (C1) is used as loop filter to design a type-II SSPLL architecture.

\subsubsection{Pseudo-differential RC Ring Oscillator}
A resistor-capacitor based RO (Fig.\ref{fig6}) has been designed with cross-coupled inverter RO units to have a linear frequency variation with respect to the control voltage ($V_C$). The RO shows a linear $K_{VCO}$ of 175 MHz/V. Pulse generator (PG) can generate equally spaced ($T_{REF}$) pulses of width 1ns to 5ns using a capacitor bank and rise/fall times have been kept equal through careful sizing.

\subsubsection{Digital Frequency Locked Loop}
Digital FLL utilizes the count of complete VCO oscillations in a user-specified number of reference clock periods to switch current sources, thereby, controlling the current flow to the RO-VCO. The multiplication factor and resolution can be controlled externally to restrict the Ring Oscillator (RO) frequency to 0.9 to 0.95 of the desired carrier frequency ($f_{c}$), which is in the locking range of ISSPLL. The FLL disengages by using switched-mode control and dead zone, enabled by dual comparators. Implemented inverter-biased current sources make the entirety of the FLL synthesis-friendly and process-portable design.
\begin{figure}
\centerline{\includegraphics[width=3.5in]{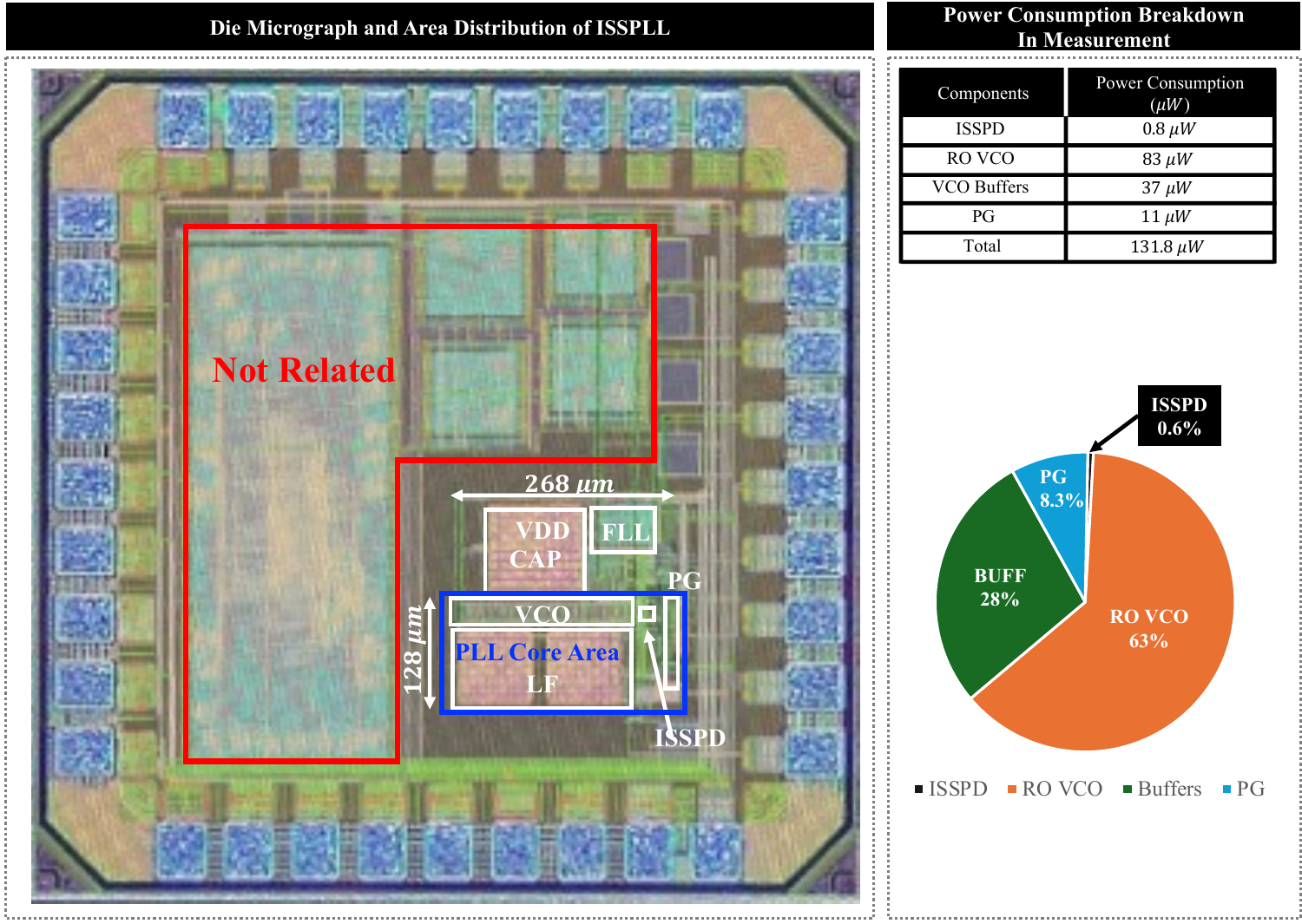}}
\caption{Area and power consumption breakdown for integrating sub-sampling PLL\label{fig7}}
\end{figure}
\vspace{-1em}\section{measurement results}
\begin{figure}
\centerline{\includegraphics[width=3.5in]{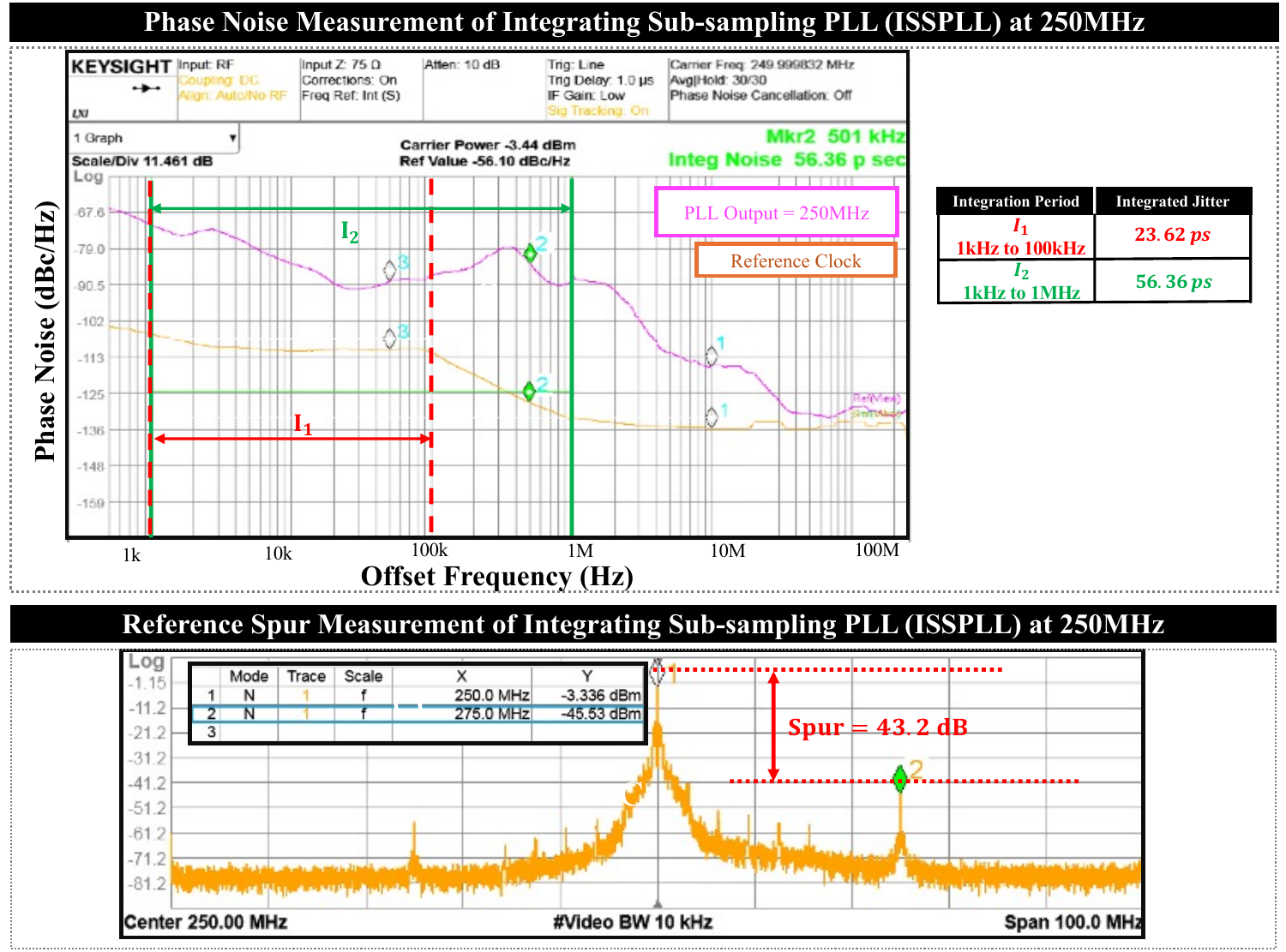}}
\caption{Measurement of Phase Noise and Reference Spur at 250MHz Output Clock Frequency\label{fig9}}
\end{figure}

The implemented ISSPLL in 65nm CMOS utilizes 0.066${mm}^2$ area for both FLL and PLL and just 0.034 ${mm}^2$ for the PLL to generate 250MHz clock from 25MHz reference clock (Fig. \ref{fig7}). The overall power consumption of ISSPLL is measured to be 131.6$\mu$W, and the power breakdown has also been shown in Fig.\ref{fig7}. The measured PN of output clock at 250MHz has been shown in Fig. \ref{fig9}. The integration range for offset frequency from 1kHz to 100kHz produces an integrated jitter of 23.62ps. The reference clock has an integrated jitter of 10.5ps over an integration range of 1kHz to 1MHz. The output spectrum (Sout) has been shown in Fig.\ref{fig9} bottom, and reference spur is observed to be -43.2dBc at 275MHz. Fig.\ref{fig7} shows the area distribution of the implemented ISSPLL, where ISSPD only occupies 5\% of the total area, thereby, reducing the total area of the ISSPLL. Overall, the area normalized jitter-power FOM ($FOM_{J,A}$) of the ISSPLL at $f_{out}$=250MHz is -236dB. 

\vspace{-1.0em}\section{conclusions and future work}
\begin{figure}
\centerline{\includegraphics[width=3.5in]{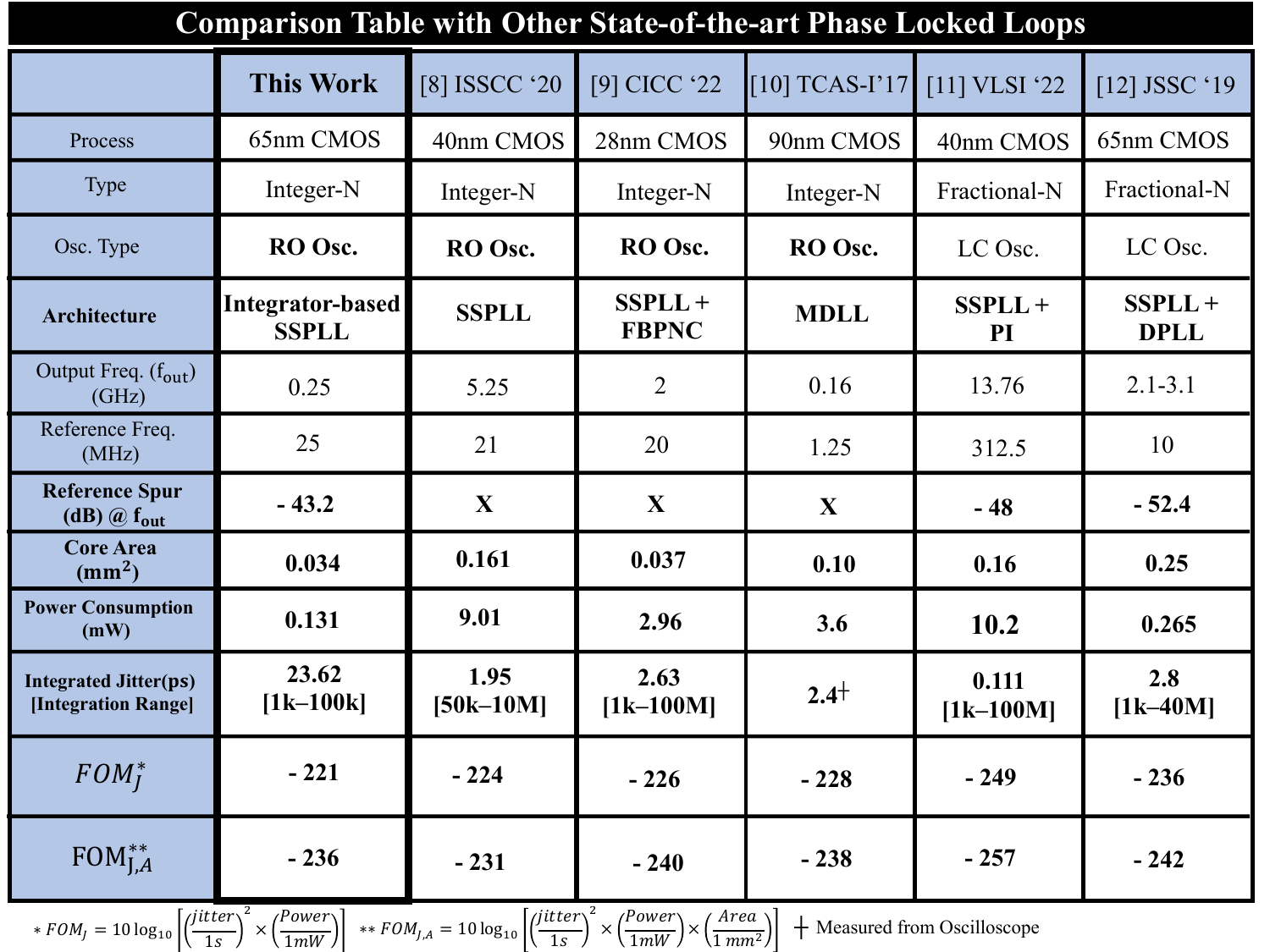}}
\caption{Comparison Table with Other State-of-the-art Phase Locked Loops 
 \label{fig11}}
\end{figure}

When compared with prior-art PLLs in Fig.\ref{fig11}, it shows improvement over RO-based SSPLL, while being comparable to LC-based SSPLLs as well. This implementation is the first SSPLL implementation that does not utilize a separate charge pump, trans-conductance amplifier, or voltage-to-current converter while improving the phase detector power consumption by 20x with respect to other state-of-the-art sub-sampling PLLs. At 250MHz, the ISSPLL shows an area normalized FOM of -236dB, while consuming a total power of 131.8$\mu W$. This will potentially motivate future integrating sub-sampling PLL architectures for higher frequency requirements as it is possible to improve the reference spur using various methodologies while utilizing the unique response of ISSPD.


 




\vfill

\end{document}